\documentclass[usenatbib]{mn2e}

\usepackage{times}
\usepackage{epsfig}
\usepackage{defs}

\def\eg{{\it e.g.}}
\def\ltsima{$\; \buildrel < \over \sim \;$}
\def\simlt{\lower.5ex\hbox{\ltsima}}   
\def\gtsima{$\; \buildrel > \over \sim \;$}
\def\simgt{\lower.5ex\hbox{\gtsima}}
\def\hide#1{}

\def\HI{\mbox{\ion{H}{1}}}

\def\GII{\mbox{\ion{He}{2}}}

\def\ie{{\it i.e.}}
\def\eg{{\it e.g.}}

\begin{document}

\title[Late Gas accretion onto Primordial Minihalos]{Late Gas accretion onto Primordial Minihalos:\\ a Model for Leo~T, Dark Galaxies 
  and Extragalactic High-Velocity Clouds}

\author[M. Ricotti]{Massimo Ricotti\thanks{E-mail: ricotti@astro.umd.edu}\\
Dept. of Astronomy, University of Maryland, College Park, MD 20742}

\maketitle

\date{\today}

\begin{abstract}
  In this letter we revisit the idea of reionization feedback on dwarf
  galaxy formation. We show that primordial minihalos with
  $v_{cir}<20$~km~s$^{-1}$ stop accreting gas after reionization, as
  it is usually assumed, but in virtue of their increasing
  concentration and the decreasing temperature of the intergalactic
  medium as redshift decreases below $z=3$, they have a late phase of
  gas accretion and possibly star formation. We expect that
  pre-reionization fossils that evolved on the outskirts of the Milky
  Way or in isolation show a bimodal star formation history with
  12~Gyr old and $<10$~Gyr old population of stars. Leo~T fits with
  this scenario. Another prediction of the model is the possible
  existence of a population of gas rich minihalos that never formed
  stars. More work is needed to understand whether a subset of compact
  high-velocity clouds can be identified as such objects or whether an
  undiscovered population exists in the voids between galaxies.
\end{abstract}
\begin{keywords}
Galaxies: formation -- Cosmology: theory
\end{keywords}

\section{Introduction}\label{sec:introduction}

It is widely assumed that reionization of the intergalactic medium
(IGM) suppresses gas accretion and galaxy formation in small mass
halos with circular velocity $v_{cir} \simlt 20$~km~s$^{-1}$,
corresponding to a virial temperature $T_{vir}\simlt 20,000$~K and to
dark halo masses $M \simlt 10^8-10^9$~M$_\odot$
\citep[\eg,][]{Babul:92,Efstathiou:92b}. Hereafter, we will refer to
these small halos affected by reionization feedback as
``minihalos''. The reason why gas accretion and star formation is
suppressed in minihalos is that, after reionization, the Jeans mass in
the IGM exceeds the mass of these minihalos and the IGM gas is unable
to condense under the influence of their gravitational
potential. Simulations of galaxy formation have confirmed this idea
\citep[\eg,][]{Bullock:00, Gnedin:00b}. However, a recently discovered
dwarf galaxy -- Leo T \citep{Ryan-Weber:08} -- despite having an
estimated mass of about $7 \times 10^6$~M$_\odot$ and having
properties typical of other ultra-faint dwarf spheroidal (dSph)
galaxies, contains gas and is actively forming stars at the present
day.  This puzzling observation prompts taking a fresh look at the
problem of gas accretion onto minihalos after reionization. However,
the model presented in this study stands independently of Leo~T.

We show that the ability of dark minihalos to accrete gas from the IGM
depends not only on the ratio of their circular velocity to the IGM
sound speed, but also on their concentration, $c$. Typically, the
concentration of a halo is $c_{vir} \sim 5$ at the redshift of
virialization but, as the halo evolves in the expanding universe, its
concentration increases. The evolution of the halo concentration with
redshift can be understood in the context of the theory of
cosmological secondary infall of dark matter \citep{Bertschinger:85}
and has been quantified using N-body simulations \citep{Bullock:01,
  WechslerB:02}.  In this letter we show that primordial minihalos
with $v_{cir}<20$~km~s$^{-1}$ stop accreting gas after reionization,
as expected, but in virtue of their increasing concentration and the
decreasing temperature of the IGM at $z<3$, they start accreting gas
from the IGM at later times. As a result, we expect that
pre-reionization fossils \citep[][hereafter RG05]{RicottiG:05} in the
Local Group have a more complex star formation history than previously
envisioned. A signature of this model is a bimodal star formation
history with an old ($\sim 12$~Gyr) and a younger ($\simlt 5-10$~Gyr,
depending on the minihalo mass) population of stars. Another
prediction of the model is the possible existence of minihalos
containing only gas but no stars. This would revive the suggestion
that some compact high-velocity clouds (CHVCs) are extragalactic
objects \citep{Blitz:99}. However, the bulk of these objects may be
yet undiscovered due to their faint 21~cm and H-$\alpha$ fluxes. The
ongoing survey ALFALFA \citep{Giovanelli:05, Giovanelli:07} may be
able to discover this new population of extragalactic objects.  This
letter is structured as follows. In \S~\ref{sec:dm} and
\S~\ref{sec:gas} we describe our model for dark matter and gas
accretion and in \S~\ref{sec:discussion} we apply the model to the
interpretation of Leo~T and observations of CHVCs. Throughout the rest
of the paper we use the following cosmological parameters ($h=0.72,
\Omega_{dm}=0.214 , \Omega_b=0.0438 , \sigma_8=0.796, n_s=0.963$) from
WMAP5 \citep{Dunkley:08}.
 
\section{Evolution of the Dark Matter Concentration}\label{sec:dm}
\begin{figure}
\centerline{\epsfig{figure=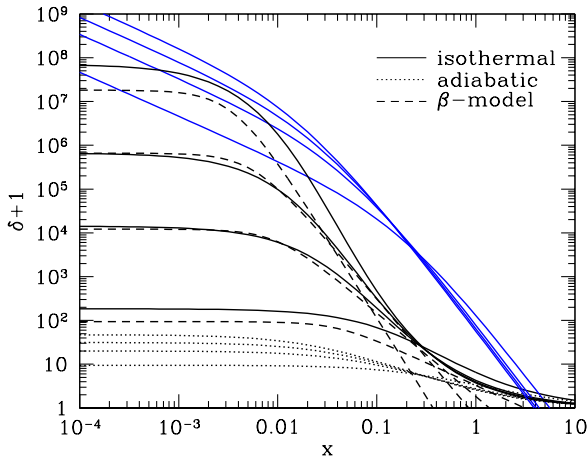,width=8cm}}
\caption{The overdensity profile of gas in hydrostatic equilibrium in
  a NFW dark matter halo virialized at $z_{vir}=0, 3, 6$ and $10$,
  (corresponding to halo concentrations, $c=5, 20, 35$ and $50$), from
  bottom to top respectively. The solid and dotted curves are for
  isothermal and adiabatic equations of state, respectively. The
  $\beta$-model provides a good fit to the isothermal profile by
  imposing the same gas density in the core.  Here, the dashed lines
  show the $beta$-model renormalized by imposing that the mean gas
  overdensity in the halo is $\Delta_g=5$ (see appendix). All halos
  have the same $\Gamma=T_{vir}/T_{igm}=1$. The halo radii are
  normalized to unity.}\label{fig:beta}
\end{figure}

N-body simulations show that the concentration of dark halos is, on
average, $c(z) \propto (1+z)^{-1}$ \citep{Bullock:01,
  WechslerB:02}. Assuming a universal NFW density profile, the halo
concentration at virialization, $c_{vir}$, has a weak dependence on
the halo mass ($\propto M^{-1/6}$). If, instead, the halo profile has
an inner slope that depends on the halos mass \citep{Ricotti:03,
  RicottiW:04, RicottiPV:07}, $c_{vir}$ is a universal
constant. Either way, although with some scatter and a dependence on
the environment, the mean concentration of minihalos is
$c(z)=c_{vir}(1+z_{vir})/(1+z)$, where $c_{vir} \sim 5$ and $z_{vir}$
depends on $v_{cir}$ and the cosmology.  Using five-year WMAP
cosmological parameters we have
\begin{eqnarray}
  v_{cir} &\sim& (17~{\rm km~s}^{-1})\left({M(z_{vir})
      \over 10^8~{\rm M}_{\odot}}\right)^{1\over 3}\left({1+z_{vir} \over
      10}\right)^{1 \over 2},\\
  \left({1+z_{vir} \over 10}\right) &\sim& {n\over 3}{\sigma_8 \over 0.79}\left[1-0.2 \log{\left({M(z_{vir}) \over 10^8~M_\odot}\right)}\right],
\end{eqnarray}
where $\sigma_8$ is the variance of density perturbations at $z=0$ in
a sphere of 8~Mpc radius, and $n$ is a number that expresses how rare
is the initial density perturbation that produces the minihalo (\ie,
the number, $n\sigma$, of standard deviations from the mean).  A halo
virialized from the collapse of an $n\sigma$ perturbation has a
typical value of the concentration $c(z,M) \approx 50
(n/3)[1-\log{(M(z_{vir})/10^8~M_\odot)^{-1/5}}]/(1+z)$. In the rest of
this letter we will use this equation as a rough estimate for the
concentrations of minihalos as a function of their mass and
redshift. This expression does not take into account the dependences
of the concentration on the local overdensity, but, qualitatively, the
model is independent of the assumed value of the concentration.
N-body simulations also show that present-day isolated dwarf halos
have a high median concentration of $c \sim 35$
\citep[\eg,][]{Colin:04}, in agreement with the extrapolation of the
fitting formula $c(z) \propto (1+z_f)/(1+z)$, derived for more massive
halos \citep[\eg,][]{Bullock:01}. However, high-resolution simulations
able to resolve the smallest minihalos (that are expected to have the
highest concentrations) typically focus on Milky Way type systems and
only a few focus on isolated minihalos in the voids.

In summary, minihalos that form at high redshift and do not merge into
bigger halos until redshift $z<1-2$, are likely to achieve a large
concentration and, as a result, have a late phase of gas accretion
from the IGM.  The evolution in relative isolation of the minihalo
also ensures that it is not tidally truncated and it does not evolve
in the Warm-Hot intergalactic medium (WHIM), a phase of the IGM that
has been shock heated to millions of degrees and believed to be the
repository of most of the baryons at $z=0$. The model has observable
effects on any mini halo surviving ``undigested'' by a larger galaxy
until $z \simlt 1-2$ (\eg, Fig~4 shows results for 2-$\sigma$
perturbations, but similar results can be derived for 1-$\sigma$
perturbations). In particular, for relatively massive pre-ionization
fossils the model predicts a bimodal star formation history. However,
in very small mass halos significant effects are only observed in the
rare high-$\sigma$ perturbations. The model that fits Leo~T uses a
3-$\sigma$ perturbation. Future work will focus on estimating how many
objects like Leo~T may exist in the Local Group and explore whether
and how many minihalos smaller than Leo~T could be detected as
extragalactic compact high velocity clouds around the Milky Way.

\section{Cold gas accretion onto primordial minihalos}\label{sec:gas}
\begin{figure}
\centerline{\epsfig{figure=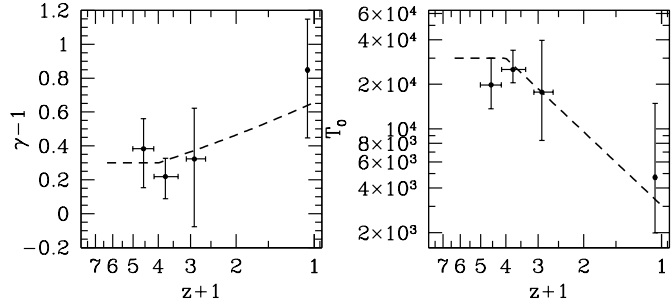,width=9cm}}
\caption{Redshift evolution of the equation of state (EOS) of the low
  density IGM from observations of Lyman-$\alpha$ forest
  \protect{\citep{RicottiGS:00}}. The dashed lines show the fiducial
  EOS used in some of our calculations. Qualitatively, our results are
  independent of the assumed fiducial model for the evolution of the
  EOS. The data point at $z=0$ shows that the low redshift evolution
  of the IGM in regions that are not shock heated (with overdensity
  less than a few) is nearly adiabatic.}
  \label{fig:eos}
\end{figure}

For a given value of the parameter $\Gamma \equiv
(v_{cir}/c^{igm}_s)^2 \equiv T_{vir}/T_{igm}$, the density profile of
the gas in a minihalo depends on its concentration and on the ability
of the gas to recombine and cool.  Figure~\ref{fig:beta} shows the gas
overdensity profile for an isothermal (solid curves) and an adiabatic
(dotted curves) gas in hydrostatic equilibrium in NFW halos with
different values of the concentration (the derivation of the equations
for the gas density profile is shown in the appendix). If the gas
condenses in the minihalo adiabatically, the overdensity in the core
does not exceed $100$. However, if the gas condenses isothermally, the
overdensity can reach values $\sim 10^7$, corresponding to gas
densities of $1-10$ cm$^{-3}$ at $z=0$.

If the cooling/heating time of the gas inside the minihalo is longer
than the Hubble time $t_H$ (that is the time scale for evolution of
the minihalo concentration and gravitational potential), the gas is
compressed and heated adiabatically (\ie, $t_{cool} \sim
t_{heat}>t_H$).  As shown in Figure~\ref{fig:eos}, after \GII~
reionization at $z \sim 3$, due to the Hubble expansion, the mean
temperature of the IGM decreases almost adiabatically because in the
low density IGM the heating time is $t_{heat} >t_H$. Similarly to the
IGM, during the initial phase of gas accretion onto a minihalo, the
gas is compressed adiabatically and the pressure prevents the gas
density in the halo core to increase substantially above the mean IGM
density. However, if the core overdensity becomes larger than a
critical value, the gas starts recombining and when the neutral
hydrogen fraction reaches a value $x_{HI} \simgt 10^{-3}$, the cooling
rate increase rapidly due to Lyman-$\alpha$ cooling. At this point
$t_{cool}<t_H$ and the gas evolution becomes nearly isothermal. The
gas overdensity in the core of a minihalo can reach the large values
shown in Figure~\ref{fig:beta} by the solid curves for an isothermal
gas.

\begin{figure}
\centerline{\epsfig{figure=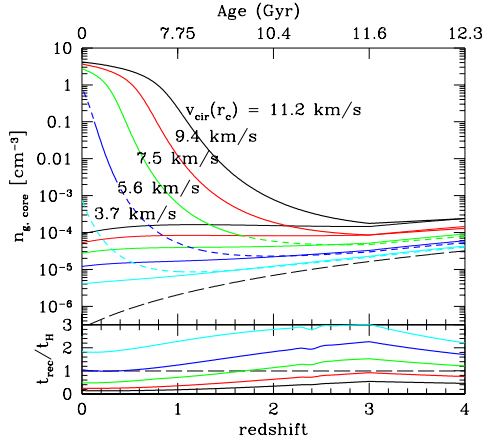,height=6.0cm}}
\caption{The evolution of the gas density in the core of dark halos
  virializing at $z_{vir}=10$ with circular velocity $v_{cir}=18, 15,
  12, 9$ and $6$~km~s$^{-1}$ (from the top to the bottom). The thick
  curves assume an isothermal gas in the halo and the thin curves
  adiabatic. The curves are dashed if the gas cannot cool
  ($t_{rec}/t_H>1$) and solid otherwise ($t_{rec}/t_H<1$). The curves
  from the bottom to the top in the lower insert panel show
  $t_{rec}/t_H$ as a function of redshift for the same halos as in the
  top panel. The EOS of the IGM is assume to evolve as in
  Fig.~2.}\label{fig:core1}
\end{figure}

The cooling time for a gas at $T \simgt 10^4$~K is always
$t_{cool}<t_{rec}$, independently of the assumed ionization fraction,
temperature and density of the gas. The cooling is due to hydrogen and
helium recombination lines if the gas is highly ionized by UV
radiation ($x_{HI} < 10^{-3}$), or hydrogen and helium Lyman-$\alpha$,
otherwise.  Thus, if the gas can recombine in a Hubble time, it is
also able to cool efficiently.  Figure~\ref{fig:core1} shows the
evolution of the core gas density (top panel) and the ratio
$t_{rec}/t_H$ (bottom panel) in minihalos of circular velocity
$v_{cir}$ for an isothermal (thick lines) and adiabatic (thin lines)
gas. Only if $t_{rec}/t_H<1$, the gas is able to collapse isothermally
(thick lines) and condense substantially in the minihalo core.

The recombination time is $t_{rec}=[n_{g, core}\alpha(T)]^{-1}$, where
$\alpha(T)$ is the hydrogen recombination rate.  Thus, at $z=0$ we
have $t_{rec}/t_H<1$ if $n_{g, core} \simgt 10^{-4}$~cm$^{-3}$.  Using
Eq.~(\ref{eq:adiab}), we also have $t_{rec}/t_H<1$ if $\beta(z) > 1.5
[(14/(1+z) - 1]$. Writing $\beta$ as a function of $v_{cir}$ we find
that minihalos with $v_{cir}>v_{cir}^{cr}$ condense isothermally.  At
redshift $z=0$, $v_{cir}^{cr} \approx c_s^{igm} 19.5/(4.4+c/4)$. These
minihalos can be observed at 21~cm and H$\alpha$ wavelengths and would
resemble faint CHVCs.  The isothermal density profile is well
approximated by a $\beta$-model (see Fig.~\ref{fig:beta}, above) with
core radius $r_{core}\approx 0.2 r_s \approx r_{vir}/25$, where
$r_{vir}$ is the virial radius at formation. After virialization, by
definition, the halo radius at redshift $z<z_{vir}$ is $r_{halo}(z)
\equiv r_{vir}(c/c_{vir})$. The typical size of the gas core is
\begin{equation}
r_{core} \approx 140~{\rm pc}\left({v_{cir} \over 17~km/s}\right)\left({10 \over 1+z_{vir}}\right)^{1.5},
\end{equation}
although, for small values of the core overdensity, most of the gas
extends out to $r_{halo} \approx 18~{\rm
  kpc}(M/10^8~M_\odot)^{1/3}/(1+z)$. The circular velocity, $v_{cir}$,
at $r_{core}$ is
\begin{equation}
v_{cir}(r_{core}) \approx 0.66 v_{cir}(r_{vir}) \approx 0.624 v_{cir}^{max},
\label{eq:vcir_c}
\end{equation}
where $v_{cir}^{max}$ is the maximum circular velocity.
\begin{figure*}
\centerline{\epsfig{figure=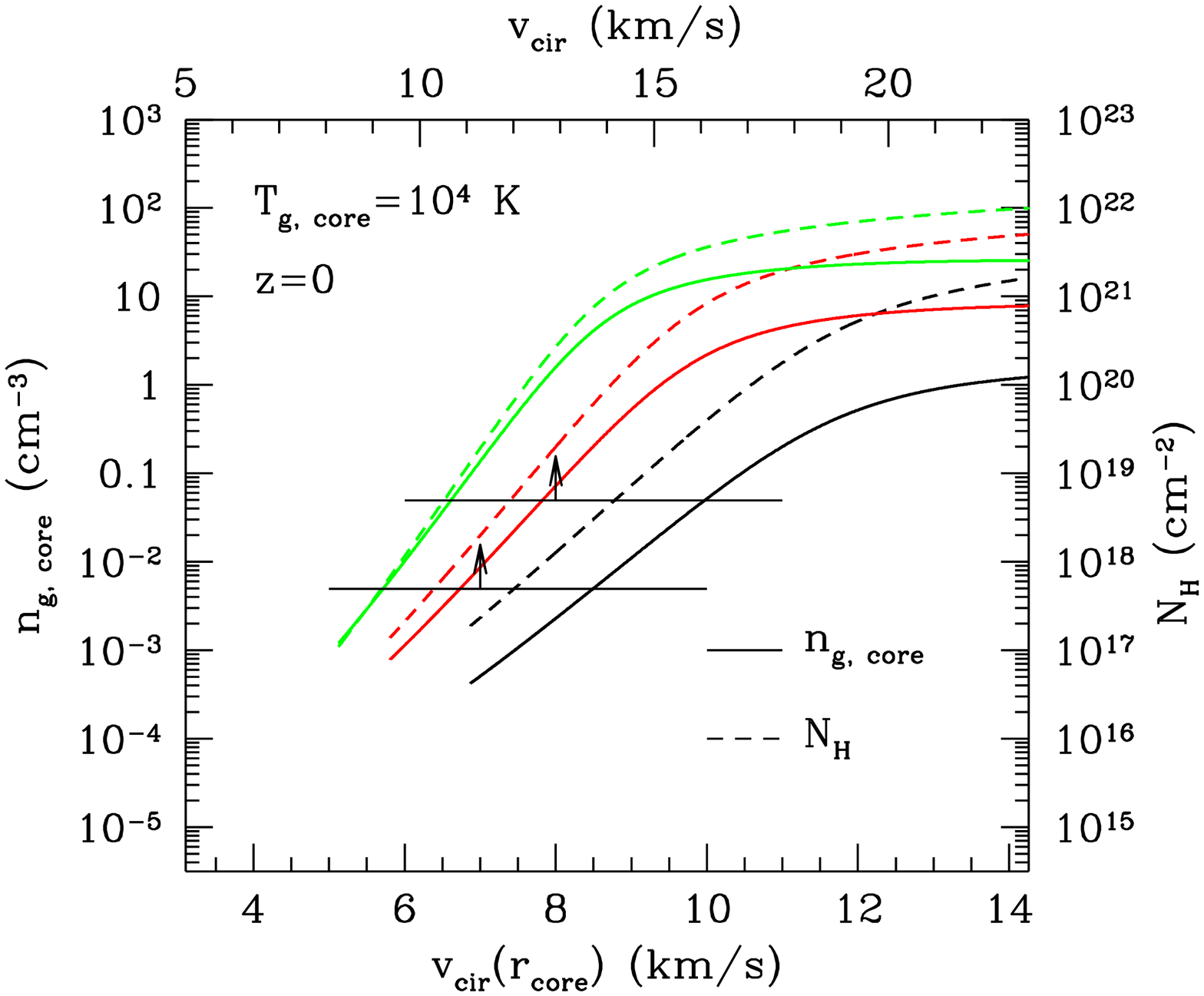,height=6.5cm}
\psfig{figure=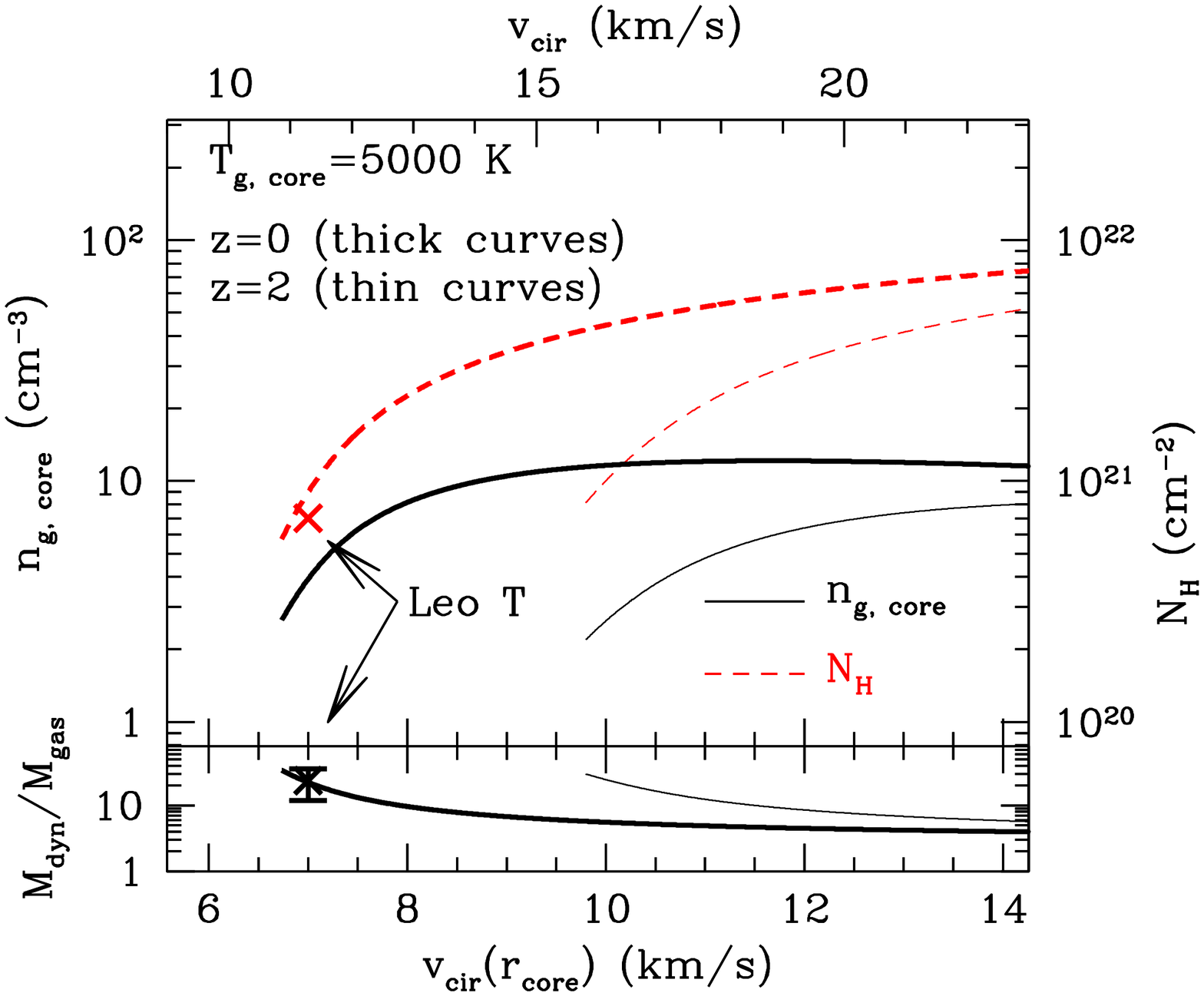,height=6.5cm}}
\caption{{\it (Left).} The gas density, $n_{g, core}$ (solid curves), and
  hydrogen column density, $N_H=2r_cn_{g, core}$ (dashed curves),
  within the core, $r_c$, of a minihalo as a function its circular
  velocity at $r_c$ at $z=0$. The curves from the bottom to the top
  refer to $2, 3$ and $4-\sigma$ perturbations. The minimum $v_{cir}$
  in each curve is determined by the condition $t_{rec}/t_H<1$,
  necessary for cooling to $T_{gas} \sim 10^4$~K. The horizontal lines
  with arrows show the requirement for cooling to temperatures below
  $10^4$~K, necessary for initiate star formation, for gas metallicity
  $Z=0.1$ (lower line) and $0.01$~Z$_\odot$ (higher line). {\it (Right).}
  Same as in the left panel but for minihalos forming from a
  3-$\sigma$ perturbation and able to form stars due to a gas
  pre-enrichment to metallicity $Z=0.1$~Z$_\odot$ and reaching a gas
  temperature $T_{g, core}=5000$~K.  The thick curves and thin curves
  refer to halos observed at redshift $z=0$ and $z=2$,
  respectively.}\label{fig:core2}
\end{figure*}

Small minihalos that can accrete gas from the IGM may not be able to
form stars because the gas cannot cool below $T\sim 10^4$~K if the gas
has low metallicity or is metal free. These halos may be stellarless
as CHVCs or may be pre-reionization fossils with a 12~Gyr old stellar
population, having gas but no young stars.  In order to form stars the
gas in the minihalo must cool below $10^4$~K and develop a multi-phase
ISM. If the gas is metal free, H$_2$ formation and cooling is unlikely
to be large enough to support star formation at $z=0$. This is
because, in a gas of primordial composition, the absence of dust grains
that typically are the main catalyst for H$_2$ formation, require that
H$_2$ forms though a chemical reaction that is very slow and that
involves the ion H$^-$ as a catalyst. At $z=0$ the flux of the H$_2$
photo-dissociating background (with energies $11.3~{\rm eV}<h
\nu<13.6$~eV) that destroys very efficiently H$_2$, is expected to be
much larger than during the dark ages at $z\sim 30$.  Positive
feedback processes that may be important at high-z in increasing the
formation rate of H$^-$ and H$_2$ are absent at $z=0$
\citep{RicottiGS:02b}. Thus, it is unlikely that dark minihalos that
were never able to form stars will be able to do that at $z=0$ for the
first time. However, due to the interesting possibility of Pop~III
formation at $z<1$ this scenario should be explored in more detail to
asses quantitatively what is the probability of Pop~III star
formation in isolated minihalos at $z=0$.

Let's now estimate which level of metal pre-enrichment is necessary
for star formation in gas rich minihalos.  The cooling function from
hyperfine transitions of oxygen and carbon depends on the gas
metallicity, $Z$, roughly as $\Lambda_{23} \sim 10^{-3}~(Z/Z_\odot)$,
where $\Lambda_{23}=10^{-23}$~erg~s$^{-1}$~cm$^{3}$. Thus, a necessary
condition for star formation is $t_{cool} \approx (0.7~{\rm
  yr})~T/(n_{g, core}\Lambda_{-23}) < t_H$, that can be written as
$n_{g, core}> n_{g, core}^{*} \approx 0.03~{\rm
  cm}^{-3}(Z/10^{-2}~Z_\odot)^{-1}$.  The left panel in
Figure~\ref{fig:core2} shows $n_{g, core}$ and $N_H$ in minihalos that
evolve isothermally at $T \sim 10^4$~K but that do not form stars
(\ie, candidates for extragalactic CHVCs). The horizontal lines shows
the requirement for metal cooling and star formation assuming gas
metallicity $Z=0.1$ and $0.01$~Z$_\odot$. The right panel in
Figure~\ref{fig:core2} shows $n_{g, core}$, $N_H$ and
$M_{dyn}/M_{gas}$ (the dynamical mass to gas mass ratio) in the core
of minihalos that are able to cool to $T=5000$~K (the temperature of
Leo~T ISM), thus, able to form stars (\ie, $n_{g, core}>n_{g,
  core}^{*}$). The symbols show the observed value for Leo~T. For
$v_{cir}(r_c)=7$~km/s, corresponding to Leo~T circular velocity, the
model gives a core radius $r_c\sim 80$~pc, that is slightly smaller
than $100$~pc measured for Leo~T. The dark matter and gas mass within
the core are $4 \times 10^6$~M$_\odot$ and $2 \times 10^5$~M$_\odot$,
respectively, in reasonable agreement with observations of Leo~T.

\section{Conclusions \& Discussion}\label{sec:discussion}

Although Leo~T contains gas, it properties as similar to those of
other ultra-faint dSph galaxies and quite different from those of more
massive dwarf irregular galaxies \citep{BovillR:08}. Its properties
are consistent with those of pre-ionization fossils but, in the
context of this model, it is difficult to understand how it could hold
on to its gas.

We point out that dark minihalos with large concentration are able to
accrete gas from the IGM more efficiently than lower concentration
halos of the same mass. The concentration of a halo depends on the
redshift of virialization as $c(z) \propto (1+z_{vir})/(1+z)$. Thus,
as a result of their growing concentration, early forming minihalos,
although affected by reionization feedback, may have a late phase of
cold gas accretion from the IGM and form stars (perhaps for the first
time) at $z<1$.  This model explains all the observed properties of
Leo~T.  Having a distance of $420$~kpc from the Milky Way, Leo~T is
just starting to fall into the Milky Way halo, thus it is likely to
have evolved in isolation (fulfilling a requirement of the model) and
unlikely to have been tidally stripped. Recent observations suggest
that Leo~T has a bimodal SFH \citep{DeJong:08} with a $>12$~Gyr
population and a $9$~Gyr population (with SF continuing until few Myrs
ago), in agreement with our expectation of a late phase of accretion
from the IGM. The bimodality of the SFH should be most pronounced in
minihalos with the lowest $v_{cir}$, gradually disappearing as
$v_{cir}$ increases toward the $ 20$~km~s$^{-1}$ threshold for
pre-reionization fossils (RG05). Leo~T has a gas velocity dispersion
$\sim 7$~km~s$^{-1}$, that, according to our model, implies a dark
halo with maximum circular velocity of $11$~km~s$^{-1}$
(Eq.~\ref{eq:vcir_c}).  Leo~T has a core radius $r_{c} \sim 100$~pc,
$M_{dyn}/M_g \sim 8-10$ \citep{SimonG:07, Ryan-Weber:08} and maximum
\HI~ column density $N_H=7\times 10^{20}$~cm$^{-2}$. These values are
all in agreement with our model as shown in \S~\ref{sec:gas} and right
panel in Fig.~\ref{fig:core2}. If future observations will show that
the dark halo of Leo~T has $v_{cir}^{max} >20$~km~s$^{-1}$ or its SFH
does not show any decline after $12$~Gyr ago, our model for Leo~T will
be disproved.

Minihalos smaller than the one hosting Leo~T can also accrete gas from
the IGM at $z<1$, but may not be able to form stars.  Sufficiently
small minihalos are likely to be dark and, if they evolved in
isolation, the IGM around them would be nearly
metal-free. Cosmological simulations show that the fraction of
primordial minihalos that host a luminous galaxy decreases steeply
with decreasing $v_{cir}$: $f(lum) \sim min[1, 50\%(v_{cir}/10~{\rm
  km/s})^{4.6}]$ \citep{RicottiGS:08}.  A fraction of dark minihalos
may be able to accrete gas from the IGM but, if the gas density in
their core is below the threshold required for star formation, these
objects will resemble CHVCs.  Additional work is needed to determine
whether our model is fully consistent with the properties of known
CHVCs \citep{Blitz:99, Braun:99, Robishaw:02}, however the typical
column densities we predict in Fig.~\ref{fig:core2}(left panel) are in
agreement with a subset of existing observations.  The typical size
and size distribution of CHVCs may agree with the model assuming a
mean distance of minihalos of 1~Mpc. We estimate that realistic
extension of the diffuse gas in dark minihalos are $\simlt
5-10$~kpc. This may exclude several known CHVC and/or set an upper
limit for their distance from the Milky Way. More extended
extragalactic CHVCs are possible for minihalo masses
$M>10^8$~M$_\odot$ that are likely to be luminous. In addition this
scenario for more massive minihalos has already been investigated by
previous studies and found not fully satisfactory \citep{Sternberg:02,
  Maloney:03, Putman:03}.  More work in this area is also motivated by
the ongoing survey ALFALFA \citep{Giovanelli:05, Giovanelli:07} that
may be able to discover a even fainter population of extragalactic
CHVCs.  Hence, future work shall focus on estimating the number, flux
and size distribution of gas rich minihalos in the Local Volume and
the local voids, producing synthetic maps for 21~cm and H-$\alpha$
wavelengths.

\appendix
\section{Gas density profile in NFW potential}

In this appendix we derive the equations describing the density
profile of intergalactic gas inside a dark minihalo. Although, the IGM
density, temperature and dark halo potential evolve with redshift, the
assumption of quasi-hydrostatic equilibrium is a good approximation
within the halo radius. Expressing the halo radius as
$r_{halo}=v_{cir} t_H/(4\Delta_{dm})^{1/2}$, where $\Delta_{dm} \sim 178$ is the
halo overdensity and $t_H$ is the Hubble time, we find that
$t_{cros}<t_H$ (where $t_{cros}=r_{halo}/c_s^{igm}$ is the halo
crossing time) if $\Gamma = (v_{cir}/c_s^{igm})^2 < (4\Delta_{dm}) \sim
800$. In this letter we are interested in halos with $\Gamma \simlt
1$, thus $t_{cros}<t_H$ and we can safely assume quasi-hydrostatic
equilibrium in order to calculate the gas density profile.
We assume a NFW density profile, $\rho \propto 1/[cx(1+cx)^2]$, with
$x \equiv r/r_{halo}$, halo concentration $c \equiv r_{halo}/r_{s}$,
where $r_s$ the core radius of the dark matter profile.  
The overdensity profile of a gas with equation of state
$P=K\rho^\gamma$, in hydrostatic equilibrium in a NFW halo of mass
$M$, concentration $c$ and circular velocity, $v_{cir}$, is
\begin{equation}
1+\delta_{g}(x)={\rho \over \rho_b}=\left[1+(\gamma-1)\ln{(1+\delta_{g}^{iso})}\right]^{1 \over \gamma-1} 
\label{eq:adiab}
\end{equation}
where $\rho_b$ is the mean IGM gas density and
$1+\delta_{g}^{iso}(x)=(1+cx)^{\beta \over c x}$ is the overdensity
for $\gamma=1$ (isothermal equation of state). In Eq.~\ref{eq:adiab}
we have normalized the density profile so that $\delta_g=0$ at
$x\rightarrow \infty$. The parameter $\beta=\Gamma c/f(c) \sim
\Gamma(4.4+0.25c)$ determines the core overdensity (at $x=0$). The gas
density profile has a core with overdensity $1+\delta_{g,
  core}=[1+(\gamma-1)\beta]^{1/(\gamma-1)}$, that for $\gamma=1$ is
$1+\delta_{g, core}^{iso}=e^\beta$.  When $\delta_{g, core}^{iso}
\simgt 10^6-10^7$, the normalization of the density profile becomes
inaccurate because the mass inside the halo exceeds the cosmic value
within the turnaround radius. In this regime the density profile must
be normalized by imposing that the total gas mass within the halo
equals a given fraction of the dark halo mass (this is necessary
because the calculation of the gas density profile does not take into
account the Hubble flow and would overestimate the gas accretion from
the IGM in this regime). The $\beta$-profile
$1+\delta_\beta(x)=\delta_c \left[1+\left({5 c
      x}\right)^2\right]^{-{3b/2}}$, where $b=0.2\beta/3$, provides a
good fit to the isothermal density profile and has the advantage of
being relatively easy to normalize to a constant mass $M_{gas}=4\pi/3
\Delta_g \rho_br_h^3$, where $\Delta_g \sim 5$ is the mean gas
overdensity inside the halo. The core density for the $\beta$-profile
with the aforementioned normalization is $\delta_c=\Delta_g
(1-b)a_1/(a_1^{(1-b)}-b)$, with $a_1=(c/0.2)^3$. We will use this
expression in our calculations of the core density for $\gamma=1$.


\bibliographystyle{./apj}
\bibliography{./leoT}

\end{document}